\documentclass[aps,prl,twocolumn,superscriptaddress]{revtex4-1}
\usepackage{graphicx}
\usepackage{latexsym}
\usepackage{amssymb}
\usepackage{amsmath}
\usepackage{amsfonts}
\usepackage{upgreek}
\usepackage{subfigure}
\usepackage{bm}
\usepackage{bbold}
\usepackage{verbatim}
\usepackage[unicode=true,
 bookmarks=false,
 breaklinks=false,pdfborder={0 0 1},backref=false,colorlinks=true]
 {hyperref}
\hypersetup{
 linkcolor=[rgb]{0,0,1},citecolor=[rgb]{0,0,1},urlcolor=[rgb]{0,0,1}}
\usepackage{multirow}
\usepackage{color}
\usepackage{comment}
\usepackage{xcolor}
\usepackage{soul}
\usepackage{tikz}

\usepackage{braket}

\begin{document}

\title{Probing a Modified Luttinger Sum Rule in the Strongly Interacting 1D Fermi-Hubbard Model}

\author{Annika B\"ohler}
\affiliation{Department of Physics and Arnold Sommerfeld Center for Theoretical Physics (ASC), Ludwig-Maximilians-Universit\"at M\"unchen, Theresienstr. 37, M\"unchen D-80333, Germany}
\affiliation{Munich Center for Quantum Science and Technology (MCQST), Schellingstr. 4, D-80799 M\"unchen, Germany}

\author{Henning Schl\"omer}
\affiliation{Department of Physics and Arnold Sommerfeld Center for Theoretical Physics (ASC), Ludwig-Maximilians-Universit\"at M\"unchen, Theresienstr. 37, M\"unchen D-80333, Germany}
\affiliation{Munich Center for Quantum Science and Technology (MCQST), Schellingstr. 4, D-80799 M\"unchen, Germany}

\author{Ulrich Schollw\"ock}
\affiliation{Department of Physics and Arnold Sommerfeld Center for Theoretical Physics (ASC), Ludwig-Maximilians-Universit\"at M\"unchen, Theresienstr. 37, M\"unchen D-80333, Germany}
\affiliation{Munich Center for Quantum Science and Technology (MCQST), Schellingstr. 4, D-80799 M\"unchen, Germany}

\author{Annabelle Bohrdt}
\affiliation{Munich Center for Quantum Science and Technology (MCQST), Schellingstr. 4, D-80799 M\"unchen, Germany}
\affiliation{University of Regensburg, Universit\"atsstr. 31, Regensburg D-93053, Germany}

\author{Fabian Grusdt}
\affiliation{Department of Physics and Arnold Sommerfeld Center for Theoretical Physics (ASC), Ludwig-Maximilians-Universit\"at M\"unchen, Theresienstr. 37, M\"unchen D-80333, Germany}
\affiliation{Munich Center for Quantum Science and Technology (MCQST), Schellingstr. 4, D-80799 M\"unchen, Germany}

\date{\today}
\begin{abstract}
Fermi surface reconstruction in cuprates can lead to an abrupt change in the Fermi momentum $k_F$ between different phases. This phenomenon remains subject of debate and is at the heart of an ongoing discussion about the nature of the metallic state in the pseudogap regime. Here we study a minimal model of a $k_F$ changing crossover in the one-dimensional Fermi-Hubbard model, where a tuning of the onsite interaction leads to a crossover between a spin-$1/2$ Luttinger liquid with small Fermi momentum and a spinless chargon liquid with large Fermi momentum. We attribute this to an emergent $U(1)$ symmetry in the strongly correlated limit, which can be used to derive a modified Luttinger sum rule recovering the large Fermi momentum. We analyse Friedel oscillations at the edge of a system to directly probe the change of Fermi momentum at zero and non-zero temperature. This paves the way for a direct experimental observation of changes of the Fermi momentum using ultracold fermions in a quantum gas microscope, with possible extensions to higher dimensional systems.
 
\end{abstract}
\maketitle

\section{Introduction}

Understanding the nature of the Fermi surface is central to understanding the electronic properties of high-$T_c$ superconductors such as cuprates. Among the many puzzling phenomena observed in these compounds is the concept of Fermi surface reconstruction in the pseudogap phase~\cite{Chowdhury_2015, Lee_2006, Norman_2005}. While a conventional large Fermi surface is found in the Fermi liquid phase of cuprates at large doping~\cite{overdoped_FS}, for the underdoped cuprates a small Fermi surface with a volume that violates the Luttinger theorem is observed~\cite{underdoped_FS}. The underlying metallic state of the pseudogap phase and the origin of the reconstructed Fermi surface remain poorly understood \cite{Chowdhury_2015}. 

Here we analyze a minimal model featuring a similar change of Fermi momentum in one dimension \cite{Soffing}. We study the one-dimensional Fermi-Hubbard (FH) model

\begin{equation}
    \hat{H}_{FH} = -t \sum_{j=0}^{L-1} \sum_{\sigma} \left (\hat{c}^\dag_{j\sigma}\hat{c}_{j+1 \sigma} + h.c.\right) + U \sum_{j=0}^{L-1} \hat{n}_{j\uparrow}\hat{n}_{j\downarrow},
\label{eq:H_FH}
\end{equation}

\noindent where $t$ denotes the hopping strength between neighbouring sites and $U$ is an onsite interaction introduced whenever two fermions occupy the same site. For large on-site interactions $U\gg t$, the low-energy physics of the FH model can be described by the related $t$-$J$ model

\begin{equation}
\begin{split}
\label{eq:H_tJ}
    \hat{H}_{tJ} = -t \hat{\mathcal{P}} \sum_{j=0}^{L-1} \sum_\sigma \left( \hat{c}_{j\sigma}^\dag \hat{c}_{j+1 \sigma} + h.c. \right)\hat{\mathcal{P}}\\
    + \frac{J}{2}\sum_{j=0}^{L-1} \left( \hat{\vec{S}}_{j} \cdot \hat{\vec{S}}_{j+1} - \frac{1}{4}\hat{n}_{j} \hat{n}_{j+1}\right),
\end{split}
\end{equation}

\noindent where $J=4t^2/U$, $\hat{n}_j=\sum_\sigma \hat{n}_{j,\sigma}$ and $\hat{\mathcal{P}}$ is a Gutzwiller projection onto maximally singly occupied sites. Note that we dropped a three-site term that appears at order $t^2/U$ \cite{Auerbach}.
For finite interaction strengths $U$, the Hubbard model of a spin-balanced system describes a Luttinger liquid of spin-1/2 particles with a Fermi momentum of $k_F = \frac{\pi}{2} n$ at filling $n=N/L$~\cite{Daul_1998}. The $U \rightarrow \infty$ limit however, where the $t$-$J$ model is valid, exhibits spin-charge separation and has been shown to be described as free spinless chargons~\cite{OgataShiba}, for which a larger Fermi momentum of $k_F = \pi n$ is found.

The goal of this paper is two-fold: On one hand we analyse the observed phenomenology in light of the Luttinger theorem by following a topological proof due to Oshikawa \cite{Oshikawa}. We show that the change in Fermi momentum can be attributed to the emergence of an additional $U(1)$ symmetry which is associated with the conservation of the number of free dopants in the $t$-$J$ model. On the other hand, we argue that these effects can be readily explored by ultracold atom experiments at currently achievable temperatures. To this end we suggest probing the crossover between the two Fermi momenta regimes by observing Friedel oscillations~\cite{Friedel, Soffing} at the boundary of a system, which can be realized by ultracold fermions in a quantum gas microscope~\cite{Bakr_2009, Cheuk_2015}. In such experiments the form of the confining potential can be engineered to observe Friedel oscillations at the boundary of a box potential or an impurity site, and the ratio of hopping $t$ and interaction strength $U$ can be tuned via the lattice potential~\cite{Bohrdt_2021, Mazurenko_2017, Bakr_2009, Cheuk_2015}. We provide simulations of the Friedel oscillations in the FH model at both zero and finite temperatures in one dimension, along with a non-interacting model in two dimensions. We suggest a possible experimental extension to the interacting case in two dimensions, and argue how observing Friedel oscillations in cold atom experiments can be a powerful tool for investigating changes in the Fermi surface of strongly interacting fermion systems.

\section{Luttinger Theorem and Emergent U(1)-Symmetry}

The Luttinger theorem \cite{Luttinger} relates the volume enclosed by the Fermi surface $V_{FS}$ of a system to the underlying particle density $n=N/L^d$ in d dimensions:

\begin{equation}
\label{eq:LT}
     \frac{V_{FS}}{(2\pi)^d} = n \mod{2\pi}.
\end{equation}

\noindent In the one-dimensional (1D) case the volume enclosed by the Fermi surface is given by $V_{FS}=2k_F$, which reduces Eq.~\eqref{eq:LT} to the statement $k_F = \pi n$. Although in one dimension Fermi surfaces are not stable and make way for Luttinger liquids, the latter are still characterized by a well-defined Fermi-momentum $k_F$ that obeys the Luttinger theorem~\cite{Luttinger_1963, Haldane_1981, Voit_1995}. Following Ref.~\cite{Oshikawa} for a topological proof of Eq.~\eqref{eq:LT}, the key idea is to adiabatically insert a $U(1)$ gauge flux quantum through a periodic model, i.e. we set $c_{j=L,\sigma}=c_{j=0,\sigma}$ in Eqs.~\eqref{eq:H_FH} and \eqref{eq:H_tJ}. Introducing this gauge flux increases the momentum in the system. Analyzing this momentum change both via a gauge transformation of the eigenstates and from an effective Fermi liquid description, one can derive Eq.~\eqref{eq:LT} \cite{Oshikawa, SachdevChowdhury, Paramekanti_2004}.

We apply the same approach to analyze the Luttinger theorem in the 1D $t$-$J$ model. Due to the Gutzwiller projection in Eq.~\eqref{eq:H_tJ}, which arises from the $U\rightarrow\infty$ limit of the hopping term in the FH model, fermions in the $t$-$J$ model are prohibited from occupying the same site even if they have opposite spins. This constitutes an additional $U(1)$ symmetry of the $t$-$J$ Hamiltonian, associated with the conservation of the total number of holes $N_h=L-\sum_{i\sigma} \hat{n}_{i\sigma}$ in the system.

\begin{figure}
    \centering
    \subfigure{\includegraphics[trim={1.2cm 0 1.2cm 0}, clip, width=0.49\columnwidth]{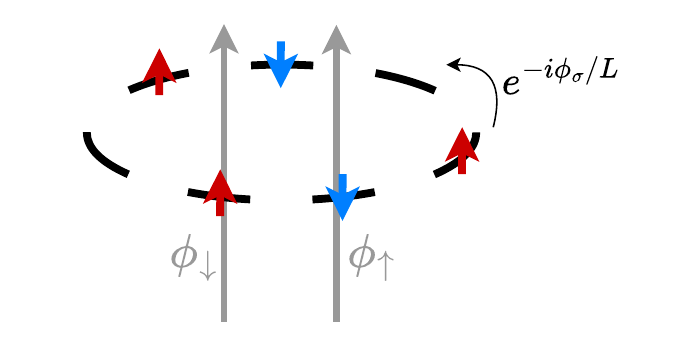}}
    \subfigure{\includegraphics[trim={1.2cm 0 1.2cm 0}, clip, width=0.49\columnwidth]{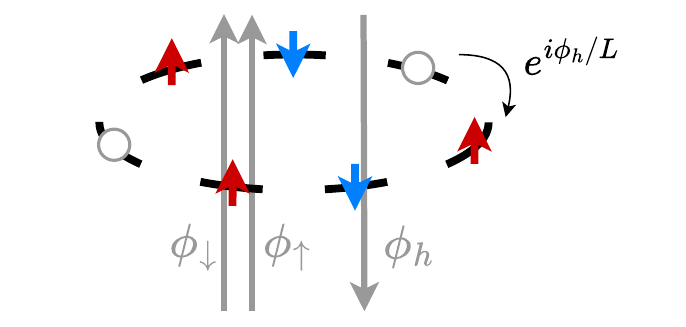}}
    \caption{Possible charge flux insertion procedures in the FH and $t$-$J$ model. Left: Fluxes $\phi_\sigma$ coupling to spin species $\sigma$ effectively modify the hopping of spinful fermions in the FH model. Right: Additional possible flux $\phi_h$ coupling to the (conserved) holes of the $t$-$J$ model.}
    \label{fig:FluxInsertion}
\end{figure}

Here, we present a modified version of the flux insertion argument, which takes into account the emergent $U(1)$ symmetry of holes explicitly in the $t$-$J$ model. To this end we note that there are different possible flux insertion procedures we can follow for the 1D $t$-$J$ and FH Hamiltonians, as summarized in Fig.~\ref{fig:FluxInsertion}. Both the $t$-$J$ and FH model exhibit two separate $U(1)$ symmetries associated with the number of particles $N_\sigma$ of each spin species $\sigma$. This allows for an insertion of $U(1)$ gauge fluxes $\phi_\sigma$, coupling to fermions of spin $\sigma$. In the case of the $t$-$J$ model an additional gauge flux $\phi_h$ coupling to the doped holes can be introduced.

In order to determine the Fermi momentum $k_F^c$ of the charges in the system, we introduce a flux $\phi_c$ coupling to all charge carriers. Note that since the holes in the $t$-$J$ model can be viewed as the charge carriers of the system, this can be achieved by considering either two equal fluxes coupling to both spin species $\phi_c=\phi_\uparrow=\phi_\downarrow$ or via a single, opposite flux coupling to the holes $\phi_c=-\phi_h$.  

\begin{figure*}[!t]
    \centering
    \includegraphics[width=\textwidth]{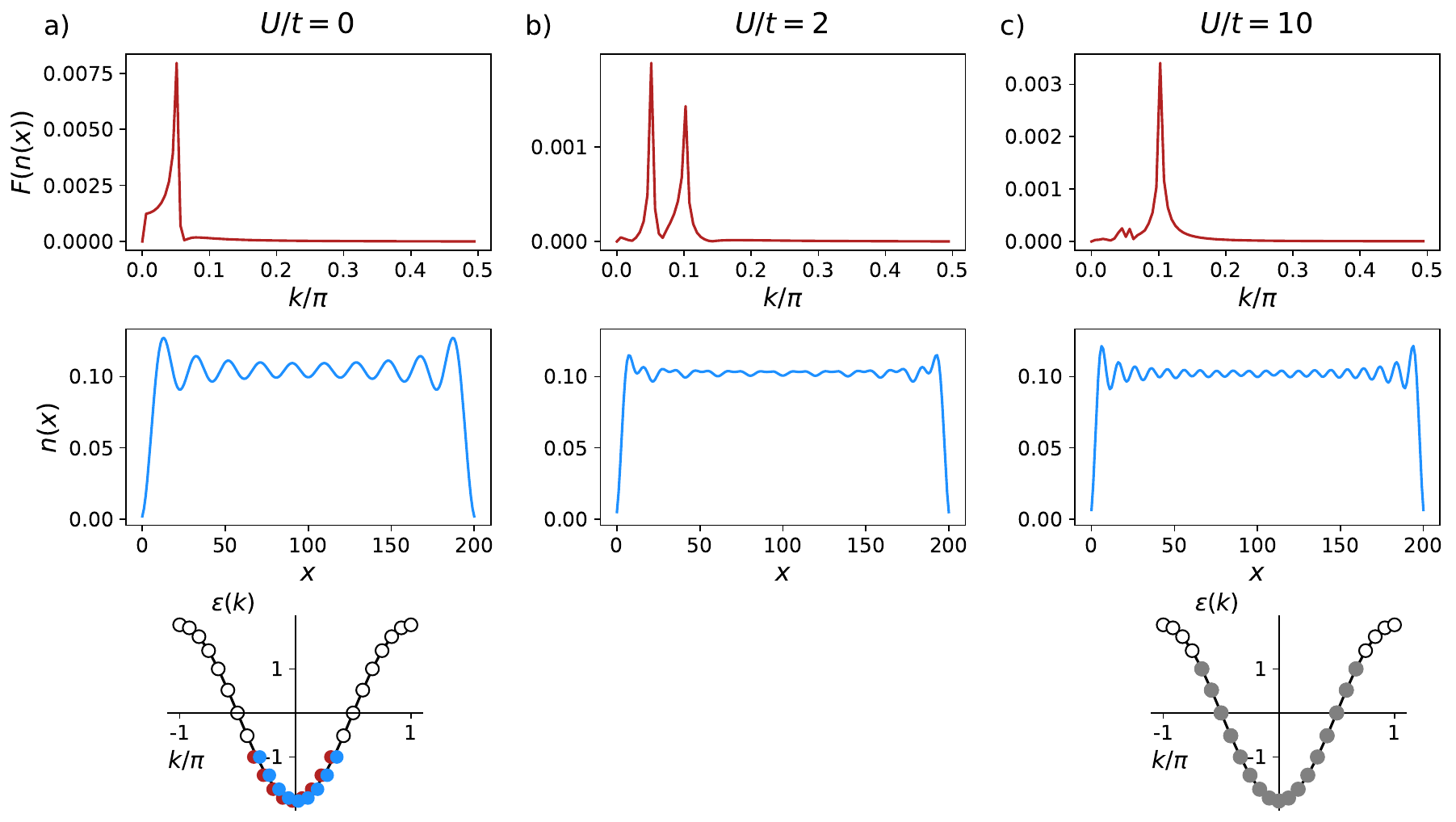}
    \caption{Zero temperature Friedel oscillations and Fourier transform for a system of size $L=200$, particle density of $n_\uparrow+n_\downarrow=0.1$, with $n_\uparrow=n_\downarrow$, and different interaction strengths $U$: (a) For $U/t=0$ a single oscillation with a frequency of $2k_F^{LL}$ corresponding to the spin-1/2 Luttinger liquid is found. (c) For $U/t=10$ the system is already close to the $t$-$J$ model limit and exhibits an approximate $U(1)$ symmetry. The oscillation frequency is found to be $2k_F^{scl}=4k_F^{LL}$, consistent with the free chargon picture, while the original Luttinger liquid frequency is strongly suppressed. (b) For intermediate interaction strengths both oscillations are present, but their amplitudes change with increasing interaction. The momentum space pictures of the respective $U=0$ and $U=\infty$ limits are illustrated in the bottom row. For the calculation of the Fourier transforms the 12 sites closest to each boundary were excluded in order to separate the effects of the decay of the oscillations towards the center of the system.}
    \label{fig:Friedel_FT}
\end{figure*}

We compare both flux insertion procedures and start by first inserting a $U(1)$ gauge flux $\phi_\sigma$ for each spin species through a periodic $t$-$J$ chain, as shown on the left-hand side of Fig.~\ref{fig:FluxInsertion}. Choosing the flux to be equal to one flux quantum $\phi_\sigma = 2\pi$, we find a new Hamiltonian $H(2\pi)$, which can be related to the original Hamiltonian $H(0)$ via a gauge transformation $\hat{H}(2\pi)=\hat{U}_\uparrow^\dag(2\pi)\hat{U}_\downarrow^\dag(2\pi)\hat{H}(0)\hat{U}_\downarrow(2\pi)\hat{U}_\uparrow(2\pi)$, where

\begin{equation}
\label{eq:gauge_trafo_spins}
    \hat{U}_{\sigma}(\varphi)=\exp\left[-i\sum_{j=1}^L (j-1)\hat{n}_{j\sigma}\frac{\varphi}{L}\right].
\end{equation}

\noindent Here $\hat{n}_{j\sigma}=\hat{c}_{j\sigma}^\dag \hat{c}_{j\sigma}$ corresponds to the number of spins $\sigma$ at site $j$, i.e. the gauge flux $\phi_\sigma$ only couples to particles with spin $\sigma$.
Inserting equal fluxes and adiabatically increasing them to one flux quantum $\phi_{\uparrow}=\phi_{\downarrow}=2\pi$ for both spin species, the ground state $\ket{\Psi_0}$ will evolve to a new state  $\ket{\Psi_\sigma}$.

In the $U\rightarrow \infty$ limit of the FH model, an additional $U(1)$ symmetry of the number of holes emerges. This becomes a full symmetry of the $t$-$J$ Hamiltonian and thus allows for an alternative flux insertion shown on the right-hand side of Fig.~\ref{fig:FluxInsertion}. This additional symmetry can be made explicit by rewriting the $t$-$J$ Hamiltonian in terms of a slave particle representation $\hat{c}_{j\sigma} = \hat{f}_{j\sigma}\hat{h}_j^\dag$, where the original particles represented by fermionic operators $\hat{c}_{j\sigma}$ are decomposed into a spinon denoted by $\hat{f}_{j\sigma}$, carrying the spin degree of freedom, and a chargon $\hat{h}_j$ carrying the charge degree of freedom~\cite{tJ_slaveparticle, spin_charge_tJ}. We choose the new flux $\phi_h$ to couple to the chargons of the system, and apply an analogous gauge transformation for $\phi_h=-2\pi$, such that $\hat{H}(2\pi)=\hat{U}_h^\dag(-2\pi) \hat{H}(0) \hat{U}_h(-2\pi)$ with

\begin{equation}
\label{eq:gauge_trafo_holes}
    \hat{U}_h(\varphi) = \exp\left[-i\sum_j (j-1) \hat{n}_{jh}\frac{\varphi}{L}\right],
\end{equation}

\noindent where $\hat{n}_{jh}=\hat{h}_j^\dag \hat{h}_j$ is the number of chargons at site $j$. Here we choose $\phi_h=-2\pi$ in order to ensure that the total flux inserted in the system is the same in both procedures and the form of the final Hamiltonian $\hat{H}(2\pi)$ will be equal in both cases. Adiabatically increasing the flux from $\phi_h=0$ to $\phi_h=-2\pi$, the ground state will evolve analogously into a new state $\ket{\Psi_h}$.

Since the system stays translationally invariant throughout the adiabatic process, $\ket{\Psi_\sigma}$ and $\ket{\Psi_h}$ must also be eigenstates of the translation operator with eigenvalues $e^{iP_\sigma}$ and $e^{iP_h}$ respectively, where $P_\sigma$ and $P_h$ are generally different from $P_0$, the momentum of the original ground state $\ket{\Psi_0}$. Expressing both the original and transformed states in the same gauge choice and evaluating $\hat{P}=\sum_{k_n,\sigma} \hat{c}_{k_n\sigma}^\dag \hat{c}_{k_n\sigma}$ we find the following changes of momenta,

\begin{equation}
\begin{split}
\label{eq:DeltaP}
    \Delta P_{\sigma} &= P_\sigma-P_0 = 2\pi \frac{N_\uparrow + N_\downarrow}{L} = 2\pi n \\
    \Delta P_{h} &= P_h-P_0= - 2\pi \frac{N_h}{L} = - 2\pi n_h,
\end{split}
\end{equation}

\noindent where $n=N/L$ is the particle density with $N=N_\uparrow + N_\downarrow$ and $n_h=N_h/L$ the hole density in the system. Making use of the fact that $\Delta P$ is only defined modulo $2\pi$ and that the number of holes is conserved such that $n_h=1-n$, we see that $- 2\pi n_h = -2\pi (1-n)=2\pi n $, proving that the two flux insertions lead to the same momentum change, consistent with our claim that they insert the same total flux into the system.

Following \cite{Oshikawa}, we continue to make use of a momentum balance argument to arrive at Eq.~\eqref{eq:LT}. To derive an alternative expression for the momentum change, we note that the flux insertion leads to an increased momentum of the quasiparticle excitations of the system, as each quasi-momentum gets shifted by $k\rightarrow k+\phi/L$~\cite{SachdevChowdhury}. After the adiabatic flux insertion this results in a shift of the entire Fermi sea by $\phi_{\sigma} /L = 2\pi/L$, or $\phi_h = -2\pi/L$ respectively, which we integrate to obtain the total momentum change $\Delta P =V_{FS}$.

We see that there are two possible flux insertion protocols for the $t$-$J$ model corresponding to the conserved $U(1)$ charges of the total spin and hole number, which we have shown lead to the same momentum change in the system, and can therefore be regarded as equivalent. However, we distinguish two possible low energy states: For a Luttinger liquid of spin-1/2 particles, there are two underlying Fermi surfaces corresponding to the two spin species. This can be represented by the momentum space picture shown on the bottom of Fig.~\ref{fig:Friedel_FT}a. We therefore obtain for the volume of the charge Fermi surface $V_{FS}^c = V_{FS}^\uparrow + V_{FS}^\downarrow$. In the free chargon picture we find that $V_{FS}^c$ directly corresponds to the Fermi surface of holes $V_{FS}^h$, and the chargon Luttinger liquid therefore has a single, larger Fermi surface, as shown on the bottom of Fig.~\ref{fig:Friedel_FT}c. As the full $U(1)$ symmetry of the total number of holes only emerges in the $t$-$J$ model, the 1D FH model at finite interaction $U$ can only be described as a spin-1/2 Luttinger liquid. However, both states are possible in the the $t$-$J$ model. Inserting $V_{FS} = 2k_F$ for the different Fermi surfaces underlying the corresponding Luttinger liquids with charge Fermi momentum $k_F^c=k_F^\uparrow=k_F^\downarrow$ and $k_F^c=\pi-k_F^h$ respectively, and comparing to the momentum change derived in Eq.~\eqref{eq:DeltaP}, we find different expressions for the two distinct scenarios:

\begin{equation}
\begin{split}
    k_F^{c,LL} &= \frac{\pi}{2} n \hspace{20pt} \text{spin-1/2 Luttinger liquid (LL)} \\
    k_F^{c, scl} &= \pi n \hspace{22pt} \text{spinless chargon liquid (scl)}
\end{split}
\end{equation}

We see that the two different low-energy states of a spin-1/2 Luttinger liquid and a spinless chargon liquid make measurably different predictions about the systems Fermi momentum in the presence of the emergent $U(1)$ symmetry. In the absence of the emergent symmetry, the only possible low-energy state is the spin-1/2 Luttinger liquid, and we conclude that the 1D FH model with finite interaction $U$ is described by a Luttinger liquid with a small Fermi momentum of $k_F=\frac{\pi}{2}n$.

\section{Signatures in Friedel Oscillations}

Since the full symmetry of the $t$-$J$ model, including the total number of holes, only emerges in the limit where $U\rightarrow\infty$ and double occupancies are completely forbidden, the question arises which perspective is more adequate for the FH Hamiltonian at large but finite $U$. To this end, we first follow Ref.~\cite{Soffing} and make use of Friedel oscillations at the edge of an open boundary system to extract the Fermi momentum $k_F$ and probe the different sum rules found above. These density oscillations are a direct result of open boundary conditions and have a frequency $f=2k_F$ proportional to the Fermi momentum~\cite{Friedel}. We also relate our observations to fluctuations of the total number of holes providing a measure for the emergent $U(1)$ symmetry.

We use density matrix renormalization group (DMRG) simulations \cite{White_1992, SchollwockDMRG2005, Schollwock_2011, hubig:_syten_toolk} of the 1D FH model in Eq.~\eqref{eq:H_FH} with open boundary conditions to extract the ground state density distributions. Fig.~\ref{fig:Friedel_FT} shows the results for the Friedel oscillations in an open boundary FH model at $L=200$ and different interaction strengths. We observe that for $U=0$ there is only a single oscillation with a frequency of $2k_F^{LL}$, where $k_F^{LL}=\pi \frac{n}{2}$ is the Fermi momentum of charges in the spin-1/2 Luttinger liquid as determined in Eq.~\eqref{eq:DeltaP}. For very large interaction strengths this is effectively replaced by an oscillation at $2k_F^{scl}=4k_F^{LL}$, which we interpret as Friedel oscillations of a spinless chargon liquid with $k_F^{scl}=2k_F^{LL} = \pi n$, see Eq.~\eqref{eq:DeltaP}. For intermediate $U$ both oscillations can be observed. As $U$ is increased and the $U(1)$ symmetry emerges, the relative amplitude of the $2k_F^{scl}$ oscillation grows, while the $2k_F^{LL}$ oscillation vanishes. The form of the two oscillations can be calculated from bosonization results~\cite{Soffing}. It has previously been shown that the crossover point where the two amplitudes are equal happens at constant $n/U$ for fixed system sizes~\cite{Soffing}.

\begin{figure}
    \centering
    \includegraphics[width=\columnwidth]{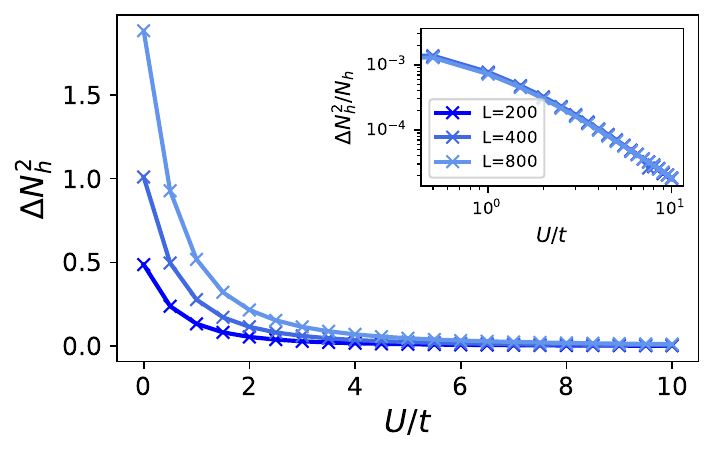}
    \caption{Hole number fluctuations $\Delta N_h ^2 = \langle N_h^2\rangle - \langle N_h \rangle^2$ for different system sizes. The fluctuations vanish with the emergence of the hole-number preserving $U(1)$ symmetry. For large $U$ the fluctuations follow a power law decay. The inset shows the relative fluctuations $\Delta N_h^2/N_h$ on a logarithmic scale. Their suppression with $U$ is independent of system size, suggesting a crossover at finite $U$ even in the thermodynamic limit.}
    \label{fig:holes}
\end{figure}

We propose to use the hole number fluctuations $\Delta N_h^2 = \langle N_h^2\rangle - \langle N_h \rangle^2$ as a good probe for the crossover between the two Fermi momenta regimes, which vanish in the case of an exact $U(1)$ conservation of holes. Fig.~\ref{fig:holes} shows $\Delta N_h^2$ for different interaction strengths $U$ and system sizes $L$, where a strong suppression of the hole number fluctuations with increasing interactions is observed as expected. The inset of Fig.~\ref{fig:holes} shows the relative fluctuations $\Delta N_h^2/N_h$ on a logarithmic scale. We find that the suppression of the relative fluctuations follows a power law for large values of $U$ and is in particular independent of system size.

\begin{figure}
    \centering
    \includegraphics[width=\columnwidth]{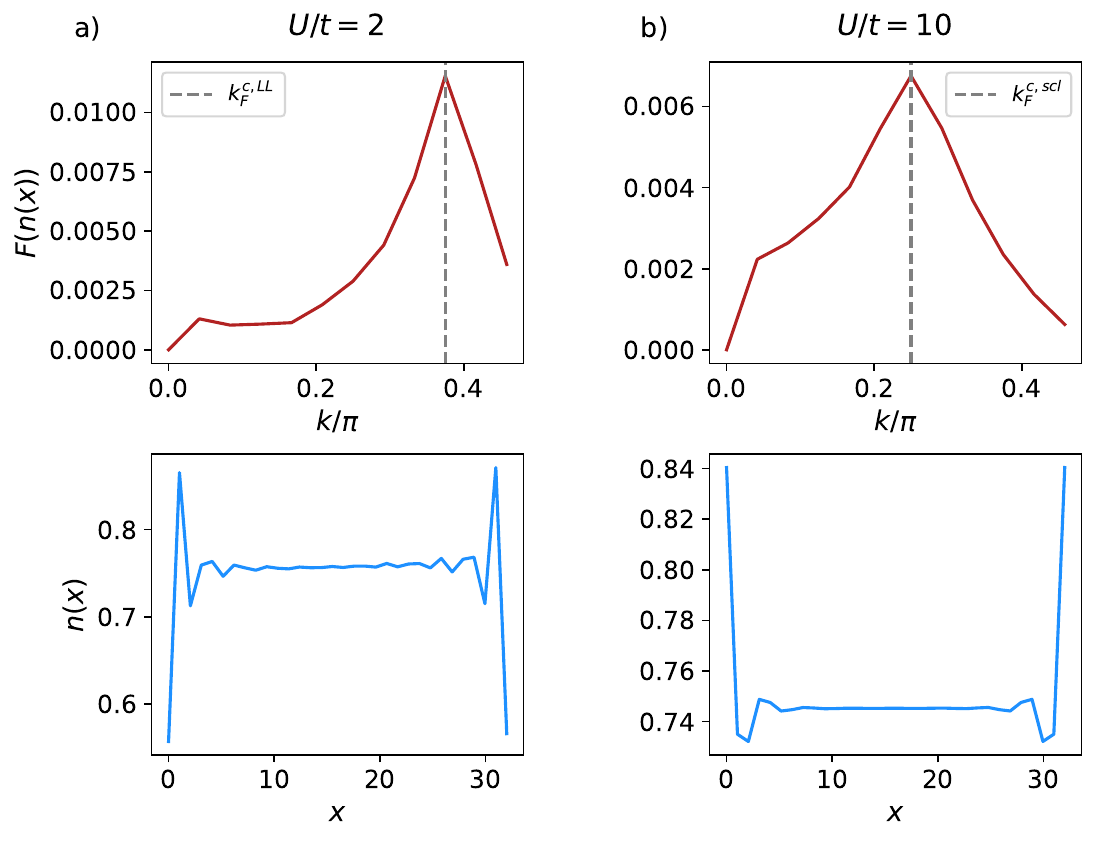}
    \caption{Finite temperature Friedel oscillations and Fourier transform for a system of $L=32$ with filling $n_\sigma=12/32$ and different interaction strengths at finite temperature $T=0.25t$, analogous to the $T=0$ case shown in Fig.~\ref{fig:Friedel_FT}. As before 4 sites close to each boundary of the system were ignored in the calculation of the Fourier transforms. Since the system is above half filling, the chargons are constituted by the double occupancies of the system, analogous to the hole-doped case studied before, changing the shape of the oscillations of the spinless chargon liquid.}
    \label{fig:Friedel_finiteT}
\end{figure}

We further simulate the Friedel oscillations in a FH model at finite temperatures in order to relate our results to realistic experimental settings. Specifically, we propose to directly measure Friedel oscillations in quantum simulators using ultracold atoms in optical lattices, see e.g. \cite{Bohrdt_2021} for a recent review. Using quantum gas microscopes with single site resolution, these systems can directly extract snapshots of the density modulations at the edges or an impurity site in the system~\cite{Bakr_2009}. Results of the finite temperature simulations in one dimension are shown in Fig.~\ref{fig:Friedel_finiteT}. We note that the thermal fluctuations introduce a faster decay of the oscillation amplitudes as one moves away from the boundary of the system, as can be seen in the real space distributions on the bottom of Fig.~\ref{fig:Friedel_finiteT}. Note that the specific form of the oscillation depends on the type of dopant. Fig.~\ref{fig:Friedel_finiteT} shows the case of $n > 1/2$, where the charge carriers conserved by the emergent symmetry are constituted by doublons, i.e. doubly occupied sites in the system. The same analysis as for the hole-doped scenario discussed above can be applied to this case, as the number of doublons also becomes conserved in the strongly correlated limit.

Our results show signatures of Friedel oscillations up to experimentally accessible temperatures of $T=0.25t$. As expected, we observe the peaks in the Fourier transform depicted in the upper panels of Fig.~\ref{fig:Friedel_finiteT} to broaden with increasing temperature, as well as a decrease in the intensity of the peaks. By tuning the filling, the frequency of the Friedel oscillations can be controlled, which allows for an observation of multiple periods even at elevated temperatures. Analysis of the peak ratio $A(k_F^\sigma)/A(k_F^c)$ suggests that a sharp crossover between the two regimes is expected to exists at finite $U>0$ even in the thermodynamic limit, which is discussed in Appendix A.

\section{Outlook - Two-Dimensional Friedel Oscillations}

\begin{figure}
    \centering
    \includegraphics[width=\columnwidth]{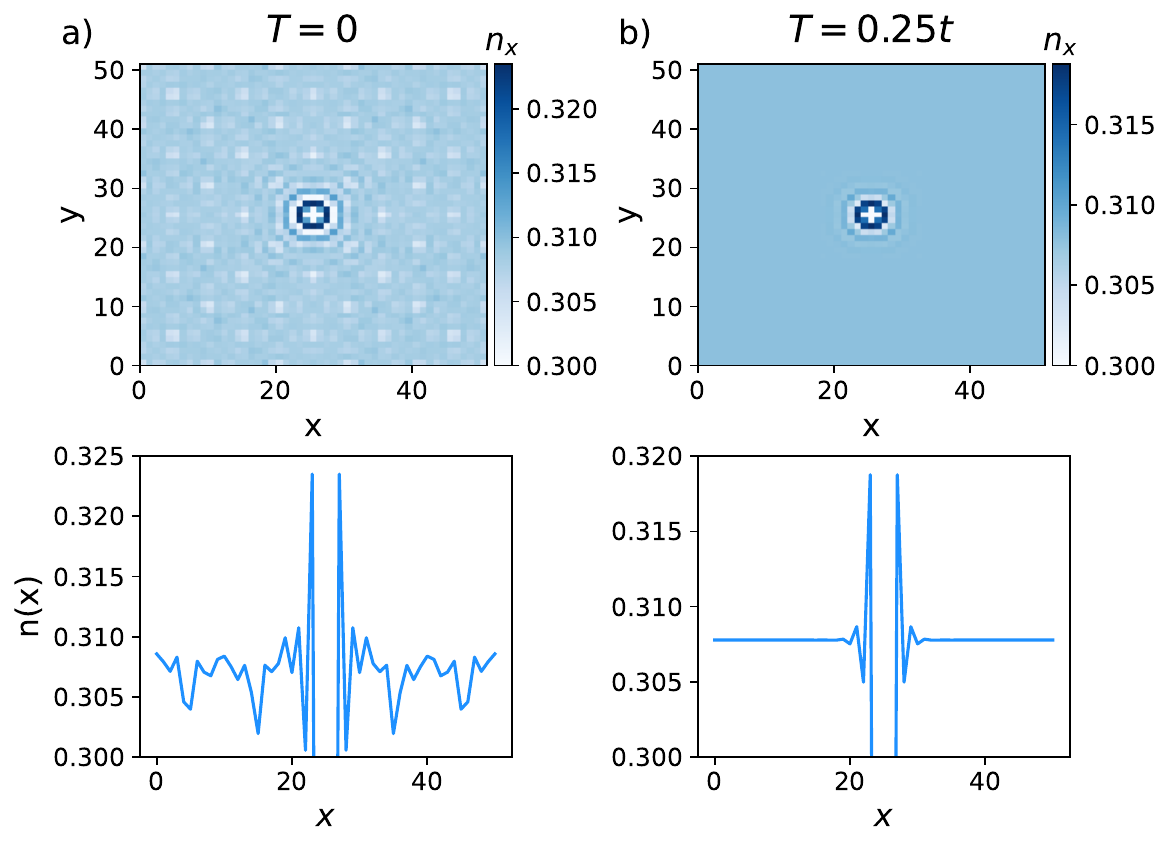}
    \caption{Friedel oscillations of non-interacting fermions, $U=0$ on a two-dimensional square lattice with filling $n_\sigma\approx0.15$ for $T=0$ (a) and $T=0.25t$ (b). The two upper panels show the real space distributions over the full lattice, below we show horizontal cuts along the impurity site. Results are obtained via exact diagonalization of a $51\times51$ lattice with an impurity site in the center and periodic boundary conditions. The impurity is added by decoupling the central site and adding a large onsite potential.}
    \label{fig:2DfiniteT}
\end{figure}

We further analyse the Hubbard model on a two-dimensional square lattice, starting with non-interacting ($U=0$) fermions. We introduce an impurity at the center of the lattice and choose periodic boundary conditions in order to prevent interference from Friedel oscillations at the boundaries. Fig.~\ref{fig:2DfiniteT} shows signatures of the Friedel oscillations for non-interacting fermions in two dimensions. The frequency of the oscillation behaves as expected from the Luttinger theorem in Eq.~\eqref{eq:LT}, where for the two-dimensional case $V_{FS}=\pi k_F^2$, which leads to a frequency of $2k_F= 4\sqrt{\pi n}$. Fig.~\ref{fig:2DfiniteT}b also shows a finite temperature calculation of the non-interacting case, allowing us to extract signatures of the Friedel oscillations up to experimentally feasible temperatures of $T=0.25t$. We see again a thermal decay of the oscillations between the zero and finite temperature calculations. A similar Fourier analysis to the one-dimensional case above allows to extract the non-interacting Fermi momentum. Extending the previous one-dimensional analysis we expect an analogous $U(1)$ symmetry to emerge in the $U\rightarrow\infty$ limit. Since numerical simulations as shown for the one-dimensional case above are much more limited in two dimensions, cold atoms could provide an interesting platform to study the large-$U$ regime of the FH model even at high doping.

\begin{figure*}
    \includegraphics[width=\textwidth]{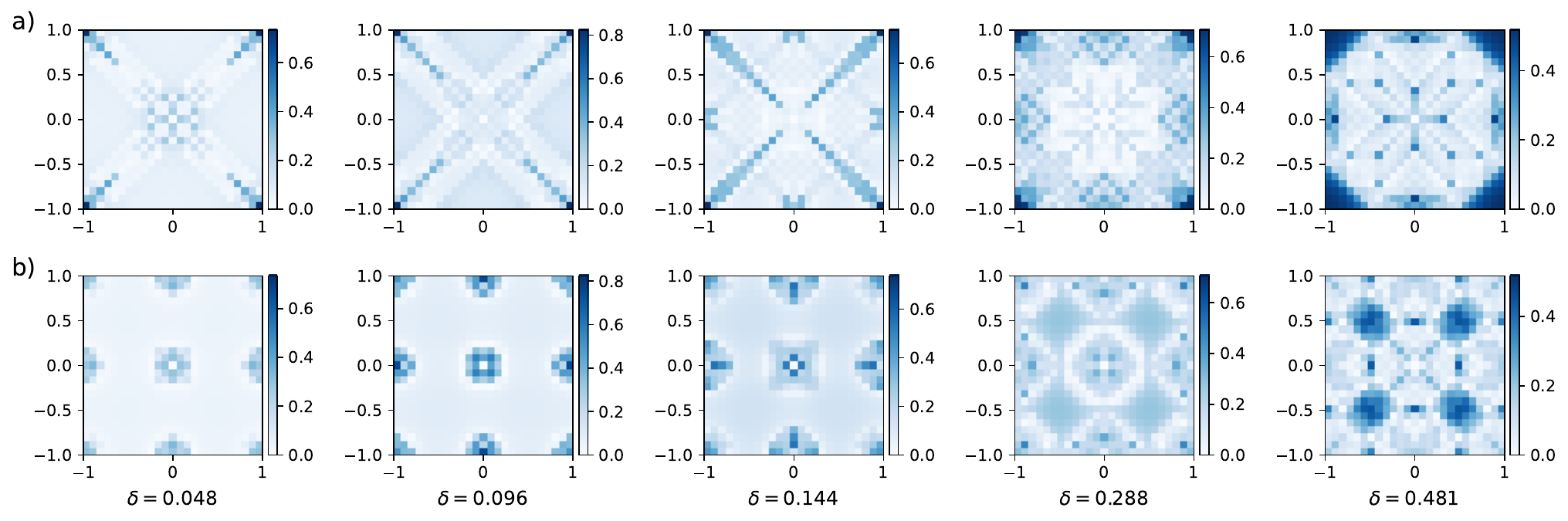}
    \caption{Fourier transform of Friedel oscillations on a square lattice of $25\times25$ sites at $T=0.001t$. Results are obtained from an exact solution of the Hamiltonians with the dispersion relations given below. The Fourier transform was taken after subtracting the mean density of the system while ignoring the central site, which is ensured to be unoccupied by adding a large onsite potential. a) Free spinful fermions with a frequency of $2k_F=\sqrt{8\pi n_\sigma}$, where $n_\uparrow=n_\downarrow=n/2$. b) 
    Magnetic polarons with a frequency of $2k_F=\sqrt{8\pi n_p}$, with $n_p=1-n$ and a smaller Brillouin zone due to the broken translational symmetry in the antiferromagnet.}
    \label{fig:polarons}
\end{figure*}

An interesting application of our analysis is the study of the formation of magnetic polarons in a 2D FH model, believed to be crucial to the pseudogap phase of cuprates. Upon doping, the FH model is expected to undergo a transition from a state resembling free spinful fermions at low filling to a system of magnetic polarons at higher fillings, associated with Fermi surface reconstruction and a change of Fermi momentum \cite{Koepsell_2021, Badoux_2016, Lee_2006}. Fig.~\ref{fig:polarons} shows a comparison of the expected Fourier signals of the corresponding density oscillations at different hole-dopings. We compare a system of free spin-1/2 fermions with a dispersion of $\epsilon(k) = -2t\cos(k)$ to free magnetic polarons in a $t$-$J$ model, which are described by the dispersion $\epsilon(k)=A(\cos(2k_x)+\cos(2k_y))+B(\cos(k_x+k_y)+\cos(k_x-k_y))$~\cite{Bohrdt_2020}. Here $A$ and $B$ depend weakly on $t/J$. For $t/J\approx2$ we take $A=0.25J$ and $B=0.36J$~\cite{polaronparams}. The signatures of both theories are strikingly different and could thus be used to distinguish the two cases in an experimental setting, where we expect that upon doping a sufficient amount of holes into the system one can observe a change from a system of magnetic polarons to a system resembling free spinful fermions. 

Friedel oscillations in clean two-dimensional cold atom systems could thus provide an alluring novel way to study the Fermi momentum and possibly even the full Fermi surface, potentially shedding new light on the nature of the Fermi surface and its reconstruction in the cuprate pseudogap phase~\cite{Cheuk_2015}, as well as a Lifshitz transition where the Fermi surface topology changes~\cite{Wu_2018}.

\section{Summary and Discussion}

We have provided a minimal model in one dimension to study a change of Fermi momentum in a strongly correlated system. To this end we have described two different pictures for the large $U$ limit of the FH model, which can be understood either in terms of a spinful Luttinger liquid or as free spinless chargons, for which different Fermi wave vectors $k_F$ are found. We have argued that the two perspectives are distinguished by an emergent $U(1)$ symmetry and have shown that this emergence drives a crossover between two regimes with different charge Fermi momenta $k_F^c$ as the Hubbard interaction is varied. We have applied a proof of the Luttinger theorem to the $t$-$J$ model as an approximation of the $U\rightarrow\infty$ limit of the FH model, making use of the full emergent symmetry in order to provide a modified flux insertion and recover the larger Fermi momentum of the spinless chargon liquid from a modified sum rule.

As the full symmetry only emerges at $U=\infty$ we propose to use Friedel oscillations in an open boundary system to probe which perspective is more adequate in strongly correlated systems. We find a smooth crossover between the two different Fermi momenta, with the Friedel oscillations of the spin-1/2 Luttinger liquid becoming strongly suppressed in the large $U$ limit. We have further extracted the hole number fluctuations at different interaction strengths and found that they are suppressed with increasing interactions, consistent with our perspective of an emergent symmetry. Finally, we have also provided a simulation of the Friedel oscillations for realistic experimental settings to extract the Fermi momentum in cold atom simulations of the full FH model in one dimension and provided an outlook onto two dimensional settings where studies of Friedel oscillations can shed light on the formation of magnetic polarons in the 2D FH model and can become an valuable tool for examining Fermi surface reconstruction believed to underlie the transition from the pseudogap to the Fermi liquid regime at high doping.

\section{Acknowledgements}
We thank Immanuel Bloch, Sebastian Eggert, Youqi Gang, Timon Hilker, Anant Kale, Lev Kendrick, Martin Lebrat, Muqing Xu and Aaron Young for fruitful discussions. We acknowledge the funding by the Deutsche Forschungsgemeinschaft (DFG, German Research Foundation) under Germany’s Excellence Strategy – EXC-2111 – 390814868. This project has received funding from the European Research Council (ERC) under the European Union’s Horizon 2020 research and innovation program (Grant Agreement no 948141) — ERC Starting Grant SimUcQuam.

\bibliography{biblio}

\appendix

\begin{figure*}
    \centering
    \includegraphics[width=\textwidth]{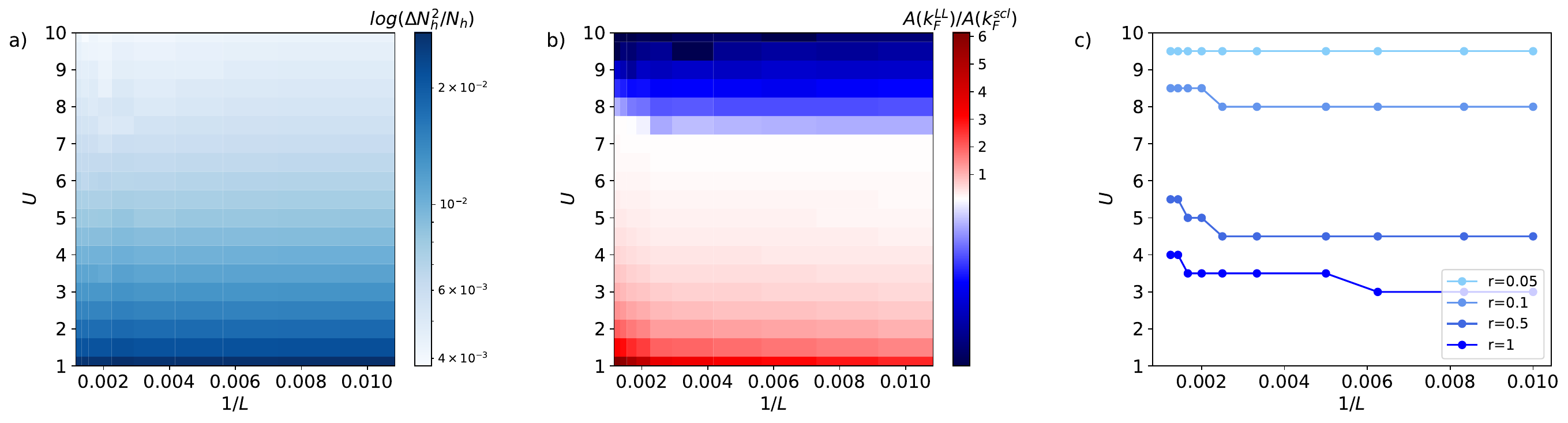}
    \caption{(a) Relative hole number fluctuations $\Delta N_h^2/N_h$ for different system sizes and interactions. The fluctuations vanish as the symmetry emerges and $N_h$ becomes a conserved quantity. This decay is independent of system size, suggesting that there is also a crossover at finite $U$ in the thermodynamic limit. (b) Ratio of Friedel oscillation amplitudes for different system sizes and interactions. The white region marks the points where $A(k_F^{LL})/A(k_F^{scl})=0.1$, where we argue that bosonization approaches for the weakly interacting limit become inaccurate. (c) System size dependence of different ratios of $A(k_F^{LL})/A(k_F^{scl})$. The interaction strength of a given ratio becomes independent of system size as the ratio is decreased, showing that weakly-interacting bosonization approaches will become inaccurate at some finite $U$ even in the thermodynamic limit.}
    \label{fig:colormaps}
\end{figure*}

\section{Thermodynamic Limit}

The question remains when and if the crossover discussed above happens in the thermodynamic limit. It has previously been suggested that for infinite system sizes the amplitude of the Friedel oscillations of the spin-1/2 Luttinger liquid would be larger than the oscillations of the spinless chargon liquid for any $U<\infty$, and therefore no crossover at finite $U$ could be observed in the thermodynamic limit
~\cite{Soffing}. However, instead of focusing on the interactions where $A(k_F^\sigma)=A(k_F^c)$, we propose as a more meaningful measure a different ratio of the two amplitudes that indicates when the system is already deep in the regime of the larger Fermi momentum. There we argue that any bosonization approach starting from the weakly correlated limit must become increasingly inaccurate, while smaller peaks at higher harmonics would still be expected even in those approaches.

Fig.~\ref{fig:colormaps}b shows the ratio of the amplitudes for different system sizes up to $L=800$. The white region corresponds to the points where $A(k_F^\sigma)/A(k_F^c)=0.1$. Here the system already exhibits an approximate $U(1)$ symmetry and conservation of the total hole number. In contrast to the $A(k_F^\sigma)=A(k_F^c)$ points, which have previously been found to depend on the system size $L$~\cite{Soffing}, the  corresponding interaction $U$ of this ratio depends less sensitively on system size.

Fig.~\ref{fig:colormaps}c shows the system size dependence of different ratios of $A(k_F^\sigma)/A(k_F^c)$. It can be seen that as this ratio is decreased, i.e. the emergent symmetry constraint becomes stronger, the corresponding $U$ becomes independent of system size.

In light of the emergent symmetry, we also argue that another, more natural quantity to measure this crossover are the relative hole number fluctuations as mentioned in the main text in Fig.~\ref{fig:holes}. Fig.~\ref{fig:colormaps}a further shows a sweep of the relative hole number fluctuations for different system sizes and interactions on a logarithmic scale, which we find to be independent of system size. Our result thus indicates that a crossover between regimes with different charge Fermi momenta happens for finite $U$ even in the thermodynamic limit.

\end{document}